\documentclass[12pt,a4paper]{article}
\usepackage{amsmath}
\usepackage{amssymb} 
\usepackage{graphicx} 
\usepackage{float}
\restylefloat{table}
\begin{document}
\begin{titlepage} 
\begin{flushright} IFUP--TH/2016\\ 
\end{flushright} ~
~
\vskip 2.5truecm 
\begin{center} 
\Large\bf Classical conformal blocks
\end{center}
\vskip 1.2truecm 
\begin{center}
{Pietro Menotti} \\ 
{\small\it Dipartimento di Fisica, Universit{\`a} di Pisa}\\ 
{\small\it 
Largo B. Pontecorvo 3, I-56127, Pisa, Italy}\\
{\small\it e-mail: pietro.menotti@unipi.it}\\ 
\end{center} 
\vskip 0.8truecm

\vskip 1 truecm

\begin{abstract}
We give a simple iterative procedure to compute the classical
conformal blocks on the sphere to all order in the modulus.
\end{abstract}

\vskip 2.0truecm

\end{titlepage}
 
\eject

\section{Introduction}

A lot of work has been  devoted about the structure of conformal
blocks in conformal field theories, and in particular in quantum Liouville
theory. There is no closed expression for such conformal blocks which
are formally defined as a power series in the modulus $x$ and there exist
recursive procedures to compute quantum conformal blocks 
(see e.g. \cite{AlZ1,AlZ2,ZZ,perlmutter,MMM,PTY}).

A remarkable conjecture has been put forward in \cite{ZZ} about the
exponentiation of such blocks in the semiclassical limit giving rise
to the so-called classical conformal blocks. Moreover such classical
conformal blocks have been shown to be connected to the accessory
parameters appearing in the auxiliary differential equation 
through a simple relation \cite{ZZ,ferraripiatek,hadaszjaskolski1}.

As a rule classical conformal blocks are obtained from the
$b\rightarrow 0$ limit of the quantum conformal blocks. In such limit
heavy cancellations occur in the quantum expression which give rise to
the above mentioned exponentiation. No general proof appears to be
available of such exponentiation process even if it has been checked
to a few orders in the expansion in $x$. 

In \cite{sphere4} the problem has been addressed of determining the
accessory parameters directly from the auxiliary equation and the
monodromy condition without appeal to the quantum conformal
blocks. The method was applied to the computation of the first two non
trivial terms in the expansion of the accessory parameter on the
sphere in the modulus $x$.

The procedure was to deform the contour in the complex plane
embracing the singularities at the origin and at the point $x$, to a
contour which embraces the cut from $1$ to infinity and closes through
a circle at infinity. The advantage of such a procedure was that one
dealt with an expansion of the energy momentum tensor which converges
for $|x|<1$ and that the asymptotic behavior of the solutions of the
accessory equation could be computed through perturbation
theory. Such perturbative series in $x$ was computed up to second
order and the result compared with success with the semiclassical
limit of the quantum conformal block.

An approach similar in nature was applied in \cite{belavin1,belavin2}
to compute 5-point classical conformal blocks for $b\rightarrow 0$
when some of the intervening charges become heavy and others stay
light. Similar limits have been considered in
\cite{hartman,perlmutter,beccaria, policastro}.

In the present paper it is shown how the method of \cite{sphere4} can
be extended to provide a very simple algebraic iterative method to
compute the accessory parameter to any order in $x$. We apply
the method explicitly to the third order in $x$ but the iterative
procedure can be carried on with great ease to any order. In fact at each
step it boils down to the solution of a linear equation.  In
\cite{ferraripiatek} and in \cite{LLNZ}, using techniques developed
in \cite{NRS}, the conformal blocks to first and second
order have been explicitly written and in
\cite{MMM} the explicit (and rather long) form of the quantum
conformal blocks has been given up to the third order included. We compare 
such values with our results finding complete agreement.

\section{The expansion of the accessory parameters}

In this section we extend the procedure of paper \cite{sphere4}
to all order in the powers of the modulus. In addition the procedure
is drastically simplified. 

To make the paper more self contained we repeat some elements
explained in \cite{sphere4}. The notation is the same as the one
adopted in \cite{sphere4}.

We start from the auxiliary differential equation of the $4$-point
Liouville problem given by
\begin{equation}\label{auxiliary}
y''(z)+Q(z)y(z)=0
\end{equation}
with
\begin{equation}
Q =\frac{\delta_0}{z^2}+ \frac{\delta}{(z-x)^2}+
\frac{\delta_1}{(z-1)^2}
+\frac{\delta_\infty-\delta_0-\delta-\delta_1}{z(z-1)}-
    \frac{C(x)}{z(z-x)(z-1)}
\end{equation}
where $\delta_j=(1-\lambda_j^2)/4$ and $C(x)$ is the accessory parameter.
The generalized monodromy problem \cite{LLNZ} is to fix the accessory 
parameter $C(x)$ by the 
requirement that the monodromy $M$ along a contour encircling both $0$ and
$x$ has a fixed trace $-2\cos\pi\lambda_\nu$. 
The idea is to use a transformation of the
variable $z$ which mimics the transformation induced by the Virasoro
generators.
Then by deforming the monodromy contour as in paper \cite{sphere4}
we find a simple iterative procedure which allow to compute the
$C(x)$ as a power expansion in $x$ to all orders.

Our $C(x)$ is related to the parameter used in \cite{ZZ,LLNZ} which we shall
call $C_L(x)$ by \break $C(x)=x(1-x)C_L(x)$. 

We supply the explicit results up to $C'''(0)$ even if it is very
simple to proceed to any order. We compare the obtained results with
the values of $C'(0)$ and $C''(0)$ given in
\cite{ferraripiatek,LLNZ} and to the expression for $C'''(0)$ derived
from \cite{MMM} finding complete agreement.

At $x=0$ the the requirement ${\rm tr}M=-2\cos\pi\lambda_\nu$ fixes
the value of $C(0)$ 

\begin{equation}
C(0)=\delta_\nu-\delta_0-\delta~.
\end{equation}
We write
\begin{equation}
Q(x) = Q_0+xQ_1+x^2Q_2+...
\end{equation}
with
\begin{equation}\label{Q0}
Q_0 = \frac{\delta_\nu}{z^2}+\frac{\delta_1}{(z-1)^2}+
    \frac{\delta_\infty-\delta_1-\delta_\nu}{z(z-1)}
\end{equation}
\begin{equation}\label{Q1}
Q_1 = \frac{2\delta-C'(0)}{z^2(z-1)}-\frac{2\delta+C(0)}{z^3(z-1)}
    \end{equation}
\begin{equation}\label{Q2}
Q_2 = -\frac{C''(0)}{2z^2(z-1)}+\frac{3\delta-C'(0)}{z^3(z-1)}-
\frac{3\delta+C(0)}{z^4(z-1)}
\end{equation}
and in general
\begin{equation}\label{Qexpansion}
Q_n=\frac{Q^{(n)}}{n!}= 
\frac{1}{z(z-1)}\bigg[\frac{-(n+1)\delta-C(0)}{z^{n+1}}+
\frac{(n+1)\delta-C'(0)}{z^n}-
\sum_{k=0}^{n-2}\frac{C^{(n-k)}(0)}{(n-k)!}\frac{1}{z^{1+k}}\bigg]~.
\end{equation}
\bigskip

The solutions of eq.(\ref{auxiliary}) for $x=0$ are known in terms of 
hypergeometric functions. 
The values of two independent solutions above the cut in $z$ 
running from $1$ to $+\infty$ are
\begin{eqnarray}\label{hypergeometric1}
y^+_1(z)&=&(1-z)^{\frac{1-\lambda_1}{2}}~z^{\frac{1-\lambda_\nu}{2}}
F(\frac{1-\lambda_1-\lambda_\infty-\lambda_\nu}{2},
\frac{1-\lambda_1+\lambda_\infty-\lambda_\nu}{2},1-\lambda_1;1-z)
\nonumber\\
&\equiv& -i e^\frac{i\pi\lambda_1}{2}t_1(z)
\end{eqnarray}
\begin{eqnarray}\label{hypergeometric2}
y^+_2(z)&=&(1-z)^{\frac{1+\lambda_1}{2}}~z^{\frac{1+\lambda_\nu}{2}}
F(\frac{1+\lambda_1+\lambda_\infty+\lambda_\nu}{2},
\frac{1+\lambda_1-\lambda_\infty+\lambda_\nu}{2},1+\lambda_1;1-z)\nonumber\\
&\equiv&-i e^{-\frac{i\pi\lambda_1}{2}} t_2(z)
\end{eqnarray}
and the asymptotic behavior at $z=+\infty+i\varepsilon$ is given by
\begin{eqnarray}\label{Y+}
Y_0^+(z)&=&
\begin{pmatrix}
-ie^\frac{i\pi\lambda_1}{2} &0\\
0 &-ie^{-\frac{i\pi\lambda_1}{2}}
\end{pmatrix}
\begin{pmatrix}
t_1(z)\\
t_2(z)
\end{pmatrix}
\equiv \Lambda_1
\begin{pmatrix}
t_1(z)\\
t_2(z)
\end{pmatrix}\nonumber\\
&\approx& \Lambda_1 B_0
\begin{pmatrix}
z^\frac{1-\lambda_\infty}{2}\\
z^\frac{1+\lambda_\infty}{2}
\end{pmatrix}
\equiv B_0^+\begin{pmatrix}
z^\frac{1-\lambda_\infty}{2}\\
z^\frac{1+\lambda_\infty}{2}
\end{pmatrix}~.
\end{eqnarray}
Similarly below the cut we have
\begin{equation}\label{Y-}
Y_0^-(z)= \Lambda_1^{-1}
\begin{pmatrix}
t_1(z)\\
t_2(z)
\end{pmatrix}
\approx 
\Lambda^{-1}_1B_0
\begin{pmatrix}
z^\frac{1-\lambda_\infty}{2}\\
z^\frac{1+\lambda_\infty}{2}
\end{pmatrix}\equiv
B^-_0
\begin{pmatrix}
z^\frac{1-\lambda_\infty}{2}\\
z^\frac{1+\lambda_\infty}{2}
\end{pmatrix}~.
\end{equation}

The explicit form of the matrix $B_0$ will not be relevant for the
following but we give it in the Appendix for
completeness.

The procedure of \cite{sphere4} is to compute the described monodromy
around $0$ and $x$ exploiting the knowledge of the asymptotic behavior
of the solution at infinity.  The chosen contour which embraces $0$
and $x$ starts at $+\infty -i\varepsilon$ i.e. below
the cut in $z$, reaches $1-i\varepsilon$ then performing a $2\pi$
clockwise rotation it reaches $1+i\varepsilon$, then it goes to
$+\infty+i\varepsilon$ and finally through a $2\pi$
anticlockwise rotation at infinity reaches the initial point
$+\infty-i\varepsilon$. It is shown in Figure 1.
\bigskip\bigskip\bigskip
\begin{figure}[htb]
\begin{center}
\includegraphics{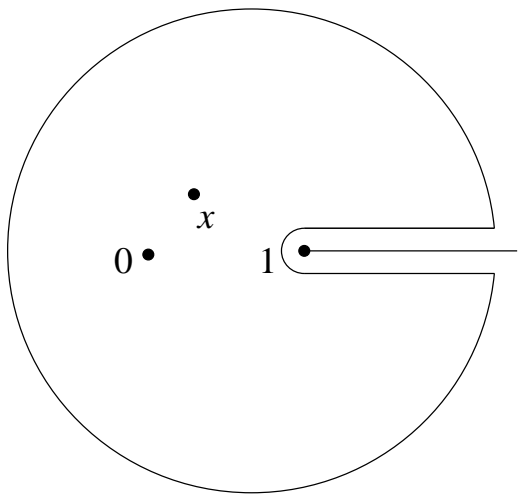}
\end{center}
\caption{The monodromy contour}
\end{figure}
The unperturbed $x=0$ monodromy matrix for such a transformation
is 
\begin{equation}
M_0= B_0^+\Lambda_\infty (B_0^-)^{-1} 
\end{equation}
where 
\begin{equation}
\Lambda_\infty=
\begin{pmatrix}
e^{i\pi(1-\lambda_\infty)}&0\\
0&e^{i\pi(1+\lambda_\infty)}
\end{pmatrix}
\end{equation}
and one easily checks \cite{sphere4} that ${\rm tr}M_0=-2\cos\pi\lambda_\nu$. 
We perform now the transformation 
\begin{equation}\label{transformation}
z(v,x)= \frac{v- {\cal C}-{\cal B}_1/v-{\cal B}_2/v^2+\dots}
{1- {\cal C}-{\cal B}_1-{\cal B}_2+\dots}
\end{equation}
where
\begin{eqnarray}
{\cal C} &=& xc_1+x^2 c_2 + x^3 c_3+\dots\nonumber\\
{\cal B}_1 &=& x^2 b_{11}+x^3 b_{12}+x^4 b_{13}+\dots\nonumber\\
{\cal B}_2 &=& x^3 b_{21}+x^4 b_{22}+x^5 b_{23}+\dots\nonumber\\
{\cal B}_3 &=& x^4 b_{31}+x^5 b_{32}+x^6 b_{33}+\dots\nonumber\\
&&.......................
\end{eqnarray} 
We note that the number of coefficients $c_k,~b_{kj}$ appearing to
order $n$ is just $n$.
Such a transformation is not one-to-one in the complex plane, but to
each finite order $n$, given some $0<r<1$, for small $x$, $z(v,x)$ 
is one-to-one for $|v|\geq r$. In fact $z(v,x)=z(v_1,x)$ for $v\neq v_1$ 
is equivalent to
\begin{equation}\label{wequation}
1 = - w w_1 \big({\cal B}_1+{\cal B}_2(w+w_1)+\dots+{\cal B}_n(w^{n-1}+
w^{n-2}w_1+\dots + w_1^{n-1})\big)
\end{equation}
where $w=1/v$, $|w|\leq 1/r$ and $w_1=1/v_1$, $|w_1|\leq 1/r$. 
On $|w|=1/r$ the r.h.s. goes
uniformly to zero for $x\rightarrow 0$ and thus for small $|x|$ by 
Rouch\'e theorem \cite{whitney} eq.(\ref{wequation}) is never
satisfied for $|w|\leq 1/r$. The monodromy contour of
Fig.1 lies in such a domain $|v|\geq r$ and from
eq.(\ref{transformation}) we have also 
$z(1,x)=1$ and $z(\infty,x)=\infty$.

Under the transformation ({\ref{transformation}) $Q_0$ goes over to
$Q_0(z(v))\big(\frac{dz}{dv}\big)^2
-\{z,v\}$ where $\{z,v\}$ is the Schwarz derivative of $z$ w.r.t. $v$
\cite{hawley}
and we shall determine $C(x)$ as to have to each order in $x$
\begin{eqnarray}\label{schwarz}
&&Q_0(z(v))\big(\frac{dz}{dv}\big)^2-\{z,v\}\nonumber\\
&=& 
\frac{\delta_0}{v^2}+\frac{\delta}{(v-x)^2}+\frac{\delta_1}{(v-1)^2}+
\frac{\delta_\infty-\delta_0-\delta-\delta_1}{v(v-1)}-
\frac{C(x)}{v(v-1)(v-x)}=Q(v)~.
\end{eqnarray}
The transformation of the solutions $Y$ is \cite{hawley}
\begin{equation}\label{Ytransformation}
Y_v(v) \equiv Y(z(v)) \bigg(\frac{dz}{dv}\bigg)^{-\frac{1}{2}}~.
\end{equation}
At infinity $z$ goes over to $(v-{\cal C})/(1-{\cal C}-{\cal B}_1-\dots)$ 
and $\displaystyle{\bigg(\frac{dz}{dv}\bigg)^{-\frac{1}{2}}}$ becomes
a constant.
Thus the power behaviors at infinity of the solutions (not the
coefficients) are unchanged i.e. they are still 
of the form $v^{\frac{1\mp \lambda_\infty}{2}}$.
Then as far as the behavior of the solution at infinity is
concerned the only
thing that changes is the matrix $B$ and thus $B^+$ and $B^-$ and 
such a change is given by the right
multiplication of $B_0$ by a diagonal matrix.

The main point in the treatment is that in computing the monodromy
matrix $M$ only the asymptotic behavior of the $Y$  
i.e. only the matrices $B^+~,B^-$ intervene. 

We shall have
\begin{equation}
B^+ = \Lambda_1 B_0 (1+ x D_1+ x^2 D_2+ \dots )
\end{equation}
\begin{equation}
B^- = \Lambda_1^{-1} B_0 (1+ x D_1+ x^2 D_2+ \dots)
\end{equation}
where $D_n$ are diagonal matrices.
It is now easily seen that the monodromy relative to
the described contour equals the one of the unperturbed (i.e. $x=0$)
case; in fact
\begin{eqnarray}
& &M = B^+\Lambda_\infty (B^-)^{-1}
=\Lambda_1 B\Lambda_\infty B^{-1}\Lambda_1\nonumber\\
&=&\Lambda_1 B_0(1+ x D_1+ x^2 D_2+ \dots)\Lambda_\infty
(1+ x D_1+ x^2 D_2+ \dots)^{-1}B_0^{-1}\Lambda_1\nonumber\\
&=&\Lambda_1 B_0\Lambda_\infty B_0^{-1}\Lambda_1=
B_0^+\Lambda_\infty (B_0^-)^{-1} =M_0~.
\end{eqnarray}
In particular the trace of the monodromy matrix for the contour embracing both
$0$ and $x$ is again $-2\cos\pi\lambda_\nu$~.

Equating the coefficients of the power expansion in $x$ of eq.(\ref{schwarz})
we find to order $n$, $n+1$ equations
corresponding to the different powers $s$ in the denominator 
\begin{equation}
\frac{1}{v^s(v-1)}~.
\end{equation}
where $s =2, \dots n+2$.
Thus we see that not only the explicit form of the matrix $B_0$ is
irrelevant but also the explicit form of the solutions i.e. the
hypergeometric functions (\ref{hypergeometric1},\ref{hypergeometric2}) 
is not relevant in the computation
but only their asymptotic behaviors (\ref{Y+},\ref{Y-}) matter.
To first order we have to fit the two terms in eq.(\ref{Q1}) with the
expansion of the l.h.s. of eq.(\ref{schwarz}) using the
two parameters $c_1,C'(0)$.
\begin{equation}
A_2
~c_1
=\begin{pmatrix}
-2\delta_\nu\\
\delta_\nu+\delta_\infty-\delta_1
\end{pmatrix}
c_1
=
\begin{pmatrix}
-2\delta-C(0)\\
2\delta -C'(0)
\end{pmatrix}
\equiv N_2
\end{equation}
i.e.
\begin{equation}
c_1 = \frac{2\delta+C(0)}{2\delta_\nu},~~~~~~~~C'(0) =
\frac{(\delta_\nu-\delta_0+\delta)(\delta_\nu-\delta_\infty+\delta_1)}{2\delta_\nu}
-C(0)
\end{equation}
The value of $c_1$ will be useful later.

To second order we have to fit the three terms in eq.(\ref{Q2}) 
finding three equations for the coefficients 
$b_{11},c_2,C''(0)$
\begin{eqnarray}\label{secondorder}
A_3\begin{pmatrix}
b_{11}\\
c_2
\end{pmatrix}&=&
\begin{pmatrix}
-3-4\delta_\nu & 0\\
3(1-\delta_1+\delta_\infty)+\delta_\nu&-2\delta_\nu\\
\delta_\nu-\delta_1-\delta_\infty&\delta_\nu+\delta_\infty-\delta_1
\end{pmatrix}
\begin{pmatrix}
b_{11}\\
c_2
\end{pmatrix}
\nonumber\\
&=&
\begin{pmatrix}
-3\delta-C(0)+3c_1^2\delta_\nu\\
c_1^2(\delta_1-2\delta_\nu-\delta_\infty)+3\delta-C'(0)\\
-C''(0)/2
\end{pmatrix}
\equiv N_3 ~.
\end{eqnarray}
As $c_1$ and $C'(0)$ are known from the previous step,
eq.(\ref{secondorder}) determines $C''(0)$ \cite{LLNZ,sphere4}
\begin{eqnarray}
& &C''(0)=-\frac{(\delta_\infty+\delta_\nu-\delta_1)
[C'(0)-3\delta+c_1^2(2\delta_\nu+\delta_\infty-\delta_1)]}{\delta_\nu}\\
&-&\frac{(C(0)+ 3\delta-3c_1^2\delta_\nu)
[3\delta_1^2+3\delta_\nu^2+3\delta_\infty(1+\delta_\infty)+
\delta_\nu(3+2\delta_\infty)-3\delta_1(1+2\delta_\nu+2\delta_\infty)]}
{\delta_\nu(3+4\delta_\nu)}~.\nonumber
\end{eqnarray}
In addition we can compute $b_{11}$ and $c_2$ which will be useful in
computing the third order.

In the Appendix we report the explicit form of the matrix $A_4$
appearing in the third order computation and the value of $C'''(0)$. 

We notice that the matrices $A_2,A_3,A_4,\dots$ are nested matrices
i.e. $A_{n+1}$ is obtained from $A_n$ by adding to the left a $n+1$
dimensional column and they are ``lower triangular'' i.e.
$A_n(h,k)=0$ for $k>h$ . This is easily seen from the nature of the
transformation (\ref{transformation}). It is trivial to carry on the
procedure to any order using eq.(\ref{Qexpansion}).
\section{Comparison with the quantum conformal blocks}
In \cite{ferraripiatek,LLNZ} the expansion of the conformal blocks up to
the second order and in \cite{MMM} the expansion up to the third order
in $x$ is given.

For the conformal blocks we have \cite{MMM}
\begin{equation}\label{quantumblock}
{\cal B}=\sum_{n=0}^\infty x^n {\cal B}^{(n)}
\end{equation}
with
\begin{equation}
{\cal B}^{(0)}=1,~~~~{\cal B}^{(1)}=\frac{(\Delta+\Delta_1-\Delta_2)
(\Delta+\Delta_3-\Delta_4)}{2\Delta} 
\end{equation}
and for the explicit expression of ${\cal B}^{(2)}$ and and the rather
long expression of ${\cal B}^{(3)}$
see \cite{MMM}. $\Delta_j$ are the quantum dimensions which are
related to the $\delta_j$ in classical limit $b\rightarrow 0$ 
by $\Delta_j = \delta_j/b^2$. 
Starting from $n=2$
the ${\cal B}^{(n)}$ also contain the central charge $c$ which for
Liouville theory is given by
\begin{equation}
c= 1+6\big(b+b^{-1}\big)^2~.
\end{equation}
The translation dictionary from the notation adopted in the present
paper and the one adopted in \cite{MMM} (which is not the same
as the one of \cite{ferraripiatek,LLNZ}) is
\begin{equation}
\delta_\nu\rightarrow \delta,~~~~\delta_0\rightarrow \delta_2,~~~~
\delta\rightarrow \delta_1,~~~~\delta_1\rightarrow \delta_3,~~~~
\delta_\infty\rightarrow \delta_4~.
\end{equation}
The relation between the conformal block and the accessory parameter 
$C_L(x)$ of \cite{ferraripiatek,LLNZ} which is related to our $C(x)$ by
$C(x)=xC_L(x)(1-x)$ is given by 
\cite{ZZ,ferraripiatek,hadaszjaskolski1,hadaszjaskolski2}
\begin{equation}
C_L(x) = \frac{\partial}{\partial x} f_\delta
\end{equation}
with 
\begin{equation}
f_\delta = (\delta-\delta_1-\delta_2) \log x
+\lim_{b\rightarrow\infty} b^2 \log \sum_{n=0}^\infty x^n {\cal B}^{(n)}
\end{equation}
defining the classical conformal blocks.
Thus we have
\begin{equation}
[xC_L(x)]|_{x=0} =\delta-\delta_1-\delta_2
\end{equation}
\begin{equation}\label{CL1}
[xC_L(x)]'|_{x=0} = \lim_{b\rightarrow 0} b^2 {\cal B}^{(1)}
\end{equation}
\begin{equation}\label{CL2}
[xC_L(x)]''|_{x=0} = \lim_{b\rightarrow 0} b^2 (4{\cal B}^{(2)}-2{{\cal B}^{(1)}}^2)
\end{equation}
\begin{equation}\label{CL3}
[xC_L(x)]'''|_{x=0} = \lim_{b\rightarrow 0} b^2~ 
3 (6{\cal B}^{(3)}-6{\cal B}^{(1)}{\cal B}^{(2)}+2{{\cal B}^{(1)}}^3)~.
\end{equation}
We note e.g. that in (\ref{quantumblock}) we have  ${\cal B}^{(3)}\sim
(\frac{1}{b^2})^3$,~ ${\cal B}^{(2)}\sim
(\frac{1}{b^2})^2$,~ ${\cal B}^{(1)}\sim
(\frac{1}{b^2})$~  
so that heavy cancellations occur in the $b\rightarrow 0$ limit in the
expressions (\ref{CL2},\ref{CL3}).
From eq.(\ref{CL1},\ref{CL2},\ref{CL3}) we can compute the
$C'(0) = [xC_L(x)]'|_{x=0}-[xC_L(x)]|_{x=0}$, 
$C''(0) = [xC_L(x)]''|_{x=0}-2[xC_L(x)]'|_{x=0}$
$C'''(0) = [xC_L(x)]'''|_{x=0}-3[xC_L(x)]''|_{x=0}$.
We checked that such values agree with the result obtained in the 
previous section and in the Appendix.

\section{Conclusions}
In the present paper we extended the technique to compute classical
conformal blocks developed in \cite{sphere4} to all order in the
modulus. The procedure is iterative and very simple. We checked the
results against the expression of the quantum conformal blocks
available in the literature finding complete agreement. One could
consider the extension of such a procedure to the computation of the
quantum corrections near the semiclassical limit.

\bigskip

\section*{Appendix}
As we mentioned in the text the explicit value of the matrix $B_0$ is
not relevant for the computations. For completeness however we report
it here below
\begin{equation}
B_0 =
\begin{pmatrix}
\frac{\Gamma(1-\lambda_1) \Gamma(-\lambda_\infty)}
{\Gamma\big(\frac{1-\lambda_1-\lambda_\infty-\lambda_\nu)}{2}\big)
\Gamma\big(\frac{1-\lambda_1-\lambda_\infty+\lambda_\nu}{2}\big)}&
\frac{\Gamma(1-\lambda_1) \Gamma(\lambda_\infty)}
{\Gamma\big(\frac{1-\lambda_1+\lambda_\infty-\lambda_\nu}{2}\big)
\Gamma\big(\frac{1-\lambda_1+\lambda_\infty+\lambda_\nu}{2}\big)}\\
\frac{\Gamma(1+\lambda_1) \Gamma(-\lambda_\infty)}
{\Gamma\big(\frac{1+\lambda_1-\lambda_\infty+\lambda_\nu)}{2}\big)
\Gamma\big(\frac{1+\lambda_1-\lambda_\infty-\lambda_\nu}{2}\big)}&
\frac{\Gamma(1+\lambda_1) \Gamma(\lambda_\infty)}
{\Gamma\big(\frac{1+\lambda_1+\lambda_\infty+\lambda_\nu}{2}\big)
\Gamma\big(\frac{1+\lambda_1+\lambda_\infty-\lambda_\nu}{2}\big)}
\end{pmatrix}~.
\end{equation}

To third order using eq.(\ref{Qexpansion}) we have the equation
\begin{equation}
\begin{pmatrix}
-12-6\delta_\nu&0&0\\
12-5\delta_1+\delta_\nu+5\delta_\infty&
-3-4\delta_\nu & 0 \\
-3\delta_1+\delta_\nu-\delta_\infty 
&3(1-\delta_1+\delta_\infty)+\delta_\nu&-2\delta_\nu\\
\delta_\nu-\delta_\infty-\delta_1
&\delta_\nu-\delta_1-\delta_\infty&\delta_\nu+\delta_\infty-\delta_1
\end{pmatrix}
\begin{pmatrix}
b_{21}\\
b_{12}\\
c_3
\end{pmatrix}
=N_4
\end{equation}
with
\begin{equation}
N_4=
\begin{pmatrix}
(10 b_{11} c_1 + 4 c_1^3) \delta_\nu-4 \delta-C(0)\\
(4 b_{11} c_1 +c_1^3)\delta_1+(-6 b_{11} c_1
-3 c_1^3 +
     6 c_1 c_2 )\delta_\nu - (4 b_{11} c_1+ c_1^3)\delta_\infty + 4 \delta -C'(0)\\
(-b_{11} c_1+2 c_1 c_2) \delta_1 -(b_{11} c_1+ 4 c_1 c_2) \delta_\nu+
      (b_{11} c_1-2 c_1 c_2) \delta_\infty - C''(0)/2\\
-C'''(0)/3!
\end{pmatrix}~.
\end{equation}

The value of $C'''(0)$ is given by
\begin{eqnarray}
& &C'''(0)=
((-C(0) - 4\delta + 10 b_{11} c_1 \delta_\nu + 4 c_1^3\delta_\nu)
   (-6 \delta_1 + 6 \delta_\nu - 6\delta_\infty))/(6 (2 +
\delta_\nu))\nonumber\\
 &+& 
 (-6 \delta_1 + 6 \delta_\nu - 6\delta_\infty)
  (-((-C(0) - 4 \delta + 10 b_{11} c_1 \delta_\nu + 4 c_1^3
\delta_\nu)\times\nonumber\\
& &   (12 - 5 \delta_1 + \delta_\nu + 5 \delta_\infty))/(6 (-3 - 4\delta_\nu)
     (2 + \delta_\nu)) \nonumber\\
&+& (C'(0) - 4 \delta - 4 b_{11} c_1 \delta_1 - 
     c_1^3 \delta_1 + 
     6 b_{11} c_1 \delta_\nu + 3 c_1^3 \delta_\nu - 6 c_1 c_2
     \delta_\nu + 4 b_{11} c_1 \delta_\infty + 
     c_1^3 \delta_\infty)/(-3 - 4 \delta_\nu))\nonumber\\ 
&+& (-6 \delta_1 + 6 \delta_\nu + 6 \delta_\infty)
  (((-C(0) - 4\delta + 10 b_{11} c_1 \delta_\nu + 4 c_1^3\delta_\nu)
     (-6 \delta_1 + 2 \delta_\nu - 2 \delta_\infty))/(24 \delta_\nu 
    (2 + \delta_\nu))\nonumber\\
 &-& (C''(0) + 2 b_{11} c_1 \delta_1 - 4 c_1 c_2 \delta_1 
+ 2 b_{11} c_1 \delta_\nu + 
     8 c_1 c_2 \delta_\nu - 2 b_{11} c_1 \delta_\infty 
  + 4c_1c_2\delta_\infty)/(4\delta_\nu)\nonumber\\ 
&+& ((6 - 6\delta_1 + 2 \delta_\nu + 6\delta_\infty)
     (-((-C(0) - 4\delta + 10 b_{11}c_1\delta_\nu + 4c_1^3\delta_\nu)\nonumber\\
& & (12 - 5\delta_1 + \delta_\nu + 5\delta_\infty))/(6(-3 - 4\delta_\nu)
        (2 + \delta_\nu))\nonumber\\ 
&+& (C'(0) - 4\delta - 4 b_{11} c_1 \delta_1 - 
      c_1^3\delta_1 + 
        6 b_{11} c_1\delta_\nu + 3 c_1^3\delta_\nu - 6 c_1c_2\delta_\nu + 
        4 b_{11}c_1\delta_\infty + c_1^3\delta_\infty)/\nonumber\\
& &(-3 - 4\delta_\nu)))/(4\delta_\nu))
\end{eqnarray}
where $C(0), C'(0), C''(0), c_1,c_2,b_{11}$ are already known from the
previous steps.

\bigskip

\section*{Acknowledgments}

The author is grateful to G. Policastro for a discussion.

\vfill


\end{document}